\documentclass[twocolumn,showpacs,showkeys,preprintnumbers,amsmath,amssymb]{revtex4}
\usepackage{graphicx}
\newcommand{\be}{\begin{equation}}
\newcommand{\ee}{\end{equation}}
\newcommand{\beq}{\begin{equation}}
\newcommand{\eeq}{\end{equation}}
\newcommand{\bea}{\begin{eqnarray}}
\newcommand{\eea}{\end{eqnarray}}
\newcommand{\eml}{\end{mathletters}}
\newcommand{\nn}{\nonumber\\}

\begin{document}

\title{Analytical approach for the Quartet Condensation Model}

\author{V.V. Baran$^{1,2}$ and D.S. Delion$^{1,3,4}$}
\affiliation{$^1$"Horia Hulubei" National Institute of Physics and 
Nuclear Engineering, \\ 
30 Reactorului, RO-077125, Bucharest-M\u agurele, Rom\^ania \\
$^2$ Department of Physics, University of Bucharest,
405 Atomi\c stilor, POB MG-11, Bucharest-M\u agurele, RO-077125, Rom\^ania \\
$^3$Academy of Romanian Scientists,
54 Splaiul Independen\c tei, RO-050094, Bucharest, Rom\^ania \\
$^4$Bioterra University, 81 G\^arlei, RO-013724, Bucharest, Rom\^ania}

\begin{abstract}
\begin{description}
\item[Background] Within the Quartet Condensation Model (QCM), the isovector pairing correlations for $N = Z$ nuclei are described with a very high accuracy by a "condensate" of $\alpha$-like quartets. The usual approach involves cumbersome recurrence relations in order to compute numerically the relevant quantities of the model: the norm of the quartet states and the mean value of the isovector pairing Hamiltonian as functions of the pair mixing amplitudes. 
\item[Purpose] We present the final analytical expressions for the above mentioned quantities, for all cases up to four quartets in the valence shell. 
\item[Method] The analytical QCM expressions were obtained by a straightforward implementation of the SO(5) algebra in the symbolic computer algebra system Cadabra2, as described below.

\item[Results] The norm of the quartet states and the mean value of the Hamiltonian are polynomial functions of the mixing amplitudes.  The numerical implementation of the QCM model is thus made trivial as a matter of copying and pasting the presented formulas. 
\item[Conclusions] We introduce in this work a method of computer aided analytical calculus for a many body setting. In particular, we provide precise and easy to use tools for the description of isovector pairing correlations. 
\end{description}
\end{abstract}



\maketitle


The $\alpha$-particle is the nucleus with the largest binding energy in nature.
For this reason this structure survives as an $\alpha$-cluster,
as can be seen from the binding energy analysis of nuclei.
Therefore the $\alpha$-cluster model of the nucleus was proposed 
in the early years of the nuclear structure theory \cite{Haf38}.
The $\alpha$-like structure is hindered by the Pauli principle
and various approaches were proposed to account for it 
\cite{Flo63,Bri66,Ari71,Wil77,Gam83,Del02}. The $\alpha$-like structures
were experimentally evidenced in light nuclei \cite{Ike68}
and therefore they were extensively analysed in the low-lying energy
region \cite{Fre97,Fre07,Hor04,Fun09,Toh17}, as well as in 
dipole resonance area \cite{He14,Chi16}.
In medium and heavy nuclei $\alpha$-clustering can experimentally be correlated 
with the $\alpha$-decay phenomenon \cite{Del10}.
It was understood that an $\alpha$-clustering component is necessary in addition to
the single-particle basis in order to describe the absolute value
of the $\alpha$-decay width \cite{Var92,Del04}. This can be explained
by the fact that $\alpha$-particles can appear only at relative low nuclear 
densities \cite{Rop98}, a situation which is realised on the nuclear surface
of $\alpha$-decaying nuclei \cite{Del13}. 

Recently the Quartet Condensation Model (QCM) was proposed for the study of isovector pairing correlations in $N=Z$ nuclei
\cite{San12a,danielphd} and further developed in \cite{San12,Neg14,San14,San15,Neg17,Sam17,Neg18} to the 
case of isoscalar pairing and $N>Z$ nuclei. 
Here, the building blocks are not the Cooper pairs, but four-body clusters 
composed of two neutrons and two protons coupled to the isospin $T = 0$ and to the angular momentum $J = 0$. 
The standard QCM procedure uses involved recurrence relations in order to compute the norm of the quartet states 
and the mean value  of the isovector pairing Hamiltonian as functions of the pair mixing amplitudes. 
Our purpose it to give closed analytical expressions for the above mentioned quantities, 
for all cases up to four quartets in the valence shell. 


We consider the isovector pairing Hamiltonian applicable to both spherical and deformed nuclei
\beq
\label{ham}
H=\sum_{i=1}^{N_\text{lev}}\epsilon_i N_{i,0}+\sum_{\tau=0,\pm1}\sum_{i,j=1}^{N_\text{lev}}V_{ij}P^{\dagger}_{i,\tau } P_{ j,\tau}~,
\eeq
where $i,j$ denote the single particle doubly-degenerate states and $\epsilon_i$ refers to the single particle energies; a time conjugated state will be denoted by $\bar{i}$. The $N_{i,0}$ operator counts the total number of particles,
$N_{i,0}=\sum_{\tau=\pi,\nu}\left(c^\dagger_{i,\tau}c_{i,\tau}+c^\dagger_{\bar{i},\tau}c_{\bar{i},\tau}\right)$, whereas the isovector triplet of pair operators is given by 
$P^\dagger_{i,\tau}=\left[c^\dagger_{i,\tau}c^\dagger_{\bar{i},\tau}\right]^{T=1}_{S=0}$. 
Explicitely, 
$P^\dagger_{i,1}=c^\dagger_{i,\nu}c^\dagger_{\bar{i},\nu},
P^\dagger_{i,-1}=c^\dagger_{i,\pi}c^\dagger_{\bar{i},\pi},
P^\dagger_{i,0}=\dfrac{1}{\sqrt{2}}\left(c^\dagger_{i,\nu}c^\dagger_{\bar{i},\pi}+c^\dagger_{i,\pi}c^\dagger_{\bar{i},\nu}\right)$. 
The $V_{ij}$ coefficients represent the matrix elements of the pairing interaction in the $\pi\pi, \nu\nu$ and $\pi\nu$ channels.

In the following we limit ourselves to a short description of the model's features for self-consistency. Within the QCM, one first defines a set of collective $\pi\pi$, $\nu\nu$ and $\pi\nu$ Cooper pairs
\beq
\Gamma^\dagger_\tau\equiv\sum_{i=1}^{N_\text{lev}}x_i P^{\dagger}_{\tau, i}~,
\eeq
where the mixing amplitudes $x_i$ are the same in all cases due to isospin invariance. A collective quartet operator is then constructed by coupling two collective pairs to the total isospin $T=0$
\beq
Q^\dagger\equiv\left[\Gamma^\dagger\Gamma^\dagger\right]^{T=0}_{S=0}\equiv 2\Gamma^\dagger_1\Gamma^\dagger_{-1}-\big(
\Gamma^\dagger_0\big)^2
\eeq
Finally, the ground state of the Hamiltonian (\ref{ham}) is described as a "condensate" (although an actual $\alpha$ condensate appears only at low densities) of such $\alpha$-like quartets
\beq
\label{condensate}
| \Psi_{q}(x) \rangle=\big(Q^{\dagger}\big)^{q}|0\rangle~,
\eeq
where $q$ is the number of quartets. By construction, this state has a well defined particle number and isospin. Its structure is defined by the mixing amplitudes $x_i$, which are determined numerically by the minimization of the Hamiltonian expectation value, subject to the unit norm constraint, i.e.
\bea
\delta\langle \Psi_{q} (x)| H| \Psi_{q}(x)\rangle&=&0~,
\nn
\langle  \Psi_{q}(x) | \Psi_{q} (x)\rangle&=&1~.
\eea
In order to compute these quantities, the method proposed in \cite{San12, danielphd} makes use of the recurrence relations obeyed by the matrix elements of the pairing interaction in the auxiliary basis $|n_1 n_2 n_3\rangle=\Gamma_{1}^{\dagger n_1}\Gamma_{-1}^{\dagger n_2}\Gamma_{0}^{\dagger n_3}|0\rangle$ of states having a well defined number of $\pi\pi, \nu\nu$ and $\pi\nu$ pairs. The advantage of this method lies in its generality:  the same numerical code is able to compute the relevant quantities in all cases of interest. On the downside, within this framework a large number of numerical evaluation steps are required in order to obtain the values of the norm and Hamiltonian average. Taking into account the fact that the minimization procedure itself requires multiple evaluations of the functions, the code running times may be considerable, especially in the case of coupled mean-field $+$ quarteting self-consistent approaches (a possible interesting generalization of the relativistic mean field $+$ projected-BCS of Ref.\cite{Las18}). Furthermore, the recurrence relations themselves are rather involved and thus challenging to derive and to implement numerically.\\

 
 We address this issues by choosing to evaluate analytically rather than numerically the  expressions of the norm and Hamiltonian average. On the one hand, a direct numerical  implementation of the final formulas considerably shortens the code running times. On the other hand, the problem of obtaining the numerical implementation itself is made trivial, being a matter of copying and pasting the formulas (with some minor syntax modification to make them compatible with the chosen programming language).\\
 
  The basic idea of our approach is that a single run of the symbolic evaluation code for an expression renders unnecesary  the alternative of an arbitrary number of possible numerical evaluations.\\
 
 To this purpose we employ the Cadabra2 symbolic computer algebra system \cite{cdb1, cdb2,cdb3}, capable of analytically handling operations with non-commuting objects.  We have implemented the SO(5) algebra (e.g. presented in \cite{danielphd}) as a set of substitution rules which are repeatedly  used  in order to evaluate the averages of the relevant operators on the quartet states. The substitution operations are performed until convergence is achieved for the considered expression. As an illustrative schematic example of the procedure, consider an average of the single particle energy term
 \beq
 \begin{aligned}
\langle 0 | \Gamma_1 \epsilon_i N_{i,1} \Gamma^\dagger_1|0\rangle&=  \epsilon_i  \langle 0 |\Gamma_1 \left(2 x_i P^\dagger_{i,1}+ \Gamma^\dagger_1N_{i,1}\right)|0\rangle\\
&=2  \epsilon_i x_i \langle 0 |\Gamma_1  P^\dagger_{i,1}|0\rangle\\
&=2  \epsilon_i x_i \langle 0 |\left(  P^\dagger_{i,1}\Gamma_1 +x_i-x_i N_{i,1} \right)|0\rangle\\
&=2  \epsilon_i x_i^2 =2\mathcal{E}_2~,
\end{aligned} 
 \eeq
 where (some of) the substitution rules employed are derived directly from the SO(5) algebra: $N_{i,1} \Gamma^\dagger_1\rightarrow 2 x_i P^\dagger_{i,1}+ \Gamma^\dagger_1N_{i,1}$  and  $\Gamma_1  P^\dagger_{i,1}\rightarrow  P^\dagger_{i,1} \Gamma_1+x_i-x_i N_{i,1}$. Also, we use the standard vacuum annihilation conditions $N_{i,1}|0\rangle \rightarrow 0$ and  $\Gamma_{1}|0\rangle \rightarrow 0$ and the notation
 $ \epsilon_i x_i^2 \rightarrow \mathcal{E}_2$. At each step, it is also necessary to invoke the routines performing the distribution of terms and the sorting of each expression. The largest running time of our brute force implementation of the SO(5) algebra are of the order of a few tens of CPU hours for the most complicated case analyzed, that of four quartets (see Eqs. (\ref{n4})-(\ref{v4}) below). However, one needs to keep in mind that the code needs to be executed only once. Also, it is not difficult to concieve further optimizations in order to reduce
the execution times and as such to easily approach the cases of five or more quartets.

  We present below the results for the cases corresponding to a number of $q=1,2,3$ and 4 quartets in the valence space. The numerical results obtained using the expressions given below were confirmed to be identical to those obtained using the standard recurrence relations approach \cite{danielprivate}.
  
   The norms of the quartet states and the Hamiltonian averages as functions of the mixing amplitudes may be expressed as
\begin{equation}
\begin{aligned}
\langle  \Psi_{q}(x) | \Psi_{q} (x)\rangle &= \mathcal{N}_{q}(x)~,\\
\langle  \Psi_{q}(x) | H| \Psi_{q}(x) \rangle &= E_{q}(x) +v_{q}(x)~.
\end{aligned}
\end{equation}

As expected, the above mentioned quantities are polynomial functions of the mixing amplitudes of degree $4q$. It is convenient to express them in terms of the sums
\begin{equation}
\begin{aligned}
\Sigma_\alpha&=\sum_{i=1}^{N_{\text{lev}}}~ x_i^\alpha~,~~~~~\mathcal{E}_\alpha=\sum_{i=1}^{N_{\text{lev}}} ~ \epsilon_i~ x_i^\alpha~,\\
\mathcal{V}_{\alpha \beta}&=\sum_{i,j=1}^{N_{\text{lev}}}~V_{ij} ~x_i^\alpha ~x_j^\beta, ~~~\mathcal{U}_{\alpha}=\sum_{i=1}^{N_{\text{lev}}}  ~V_{ii}~ x_i^\alpha~, \\\end{aligned}
\end{equation}
\begin{widetext}
where $x_i^\alpha$ is the amplitude $x_i$ to the power $\alpha$ and ${N_{\text{lev}}}$ is the number of levels in the valence space. The  generalization to the case of degenerate (spherical) levels is made trivial by the fact that the mixing amplitudes and interaction matrix elements are equal within each degenerate subspace.\\

For $q=1$, we obtain
\begin{equation}
\begin{aligned}
\mathcal{N}_1&=3\big(2~{\Sigma_2}^2+~{\Sigma_4} \big)~,\\
E_1&=12\left(2~\mathcal{E}_2~{\Sigma_2}+~\mathcal{E}_4\right)~,\\
v_1&=3~ ( 4 ~\Sigma_2~  \mathcal{V}_{1,1}  +4~ \mathcal{V}_{1,3} +  \mathcal{U}_4 )~.\\
\end{aligned}
\end{equation}

For $q=2$ the results are
\begin{equation}
\label{q2}
\begin{aligned}
\mathcal{N}_{2}&=30 \big({\Sigma_8}+4~{\Sigma_2}^4+7~{\Sigma_4}^2-8~{\Sigma_2}~{\Sigma_6}
-4~{\Sigma_4}~{\Sigma_2}^2\big)~,\\
E_2&=240\big(
~\mathcal{E}_{8}+4~\mathcal{E}_{2} ~{\Sigma_{2}}^{3}+7~\mathcal{E}_{4} ~\Sigma_{4}-2~\mathcal{E}_{2} ~\Sigma_{6}-6~\mathcal{E}_{6} ~\Sigma_{2}-2~\mathcal{E}_{4} ~{\Sigma_{2}}^{2}-2~\mathcal{E}_{2} ~\Sigma_{2} ~\Sigma_{4}\big)~,\\
v_2&=60 \big(8 ~{\Sigma_2}^3 ~\mathcal{V}_{1,1}-8 ~{\Sigma_2}^2 ~\mathcal{V}_{1,3}-4~\Sigma_4
  ~\Sigma_2 ~\mathcal{V}_{1,1}-24~\Sigma_2 ~\mathcal{V}_{1,5}-12~\Sigma_2 ~\mathcal{V}_{3,3}-4
  ~\Sigma_6 ~\mathcal{V}_{1,1}\\&
  +28~\Sigma_4 ~\mathcal{V}_{1,3}+4 ~\mathcal{V}_{1,7}+4
  ~\mathcal{V}_{3,5}+14 ~{\Sigma_2}^2 ~\mathcal{U}_4
   +4~\Sigma_2 ~\mathcal{U}_6-9~\Sigma_4
  ~\mathcal{U}_4-~\mathcal{U}_8\big)~.\\  
\end{aligned}
\end{equation}

The expressions for $q=3$ of the norm function, single particle energy and interaction terms read
\begin{equation}
\begin{aligned}
\mathcal{N}_3&=630\big(6~{\Sigma_{12}}-24~\Sigma_{2} ~{\Sigma_{10}}+8~{\Sigma_{2}}^{6}-57~\Sigma_{4} ~\Sigma_{8}-12~\Sigma_{4} ~{\Sigma_{2}}^{4}+27~{\Sigma_{4}}^{3}+68~{\Sigma_{6}}^{2}-8~\Sigma_{6} ~{\Sigma_{2}}^{3}+52~\Sigma_{8} ~{\Sigma_{2}}^{2}\\
&+26~{\Sigma_{2}}^{2} ~\Sigma_{8}+22~{\Sigma_{2}}^{2} ~{\Sigma_{4}}^{2}-120~\Sigma_{2} ~\Sigma_{4} ~\Sigma_{6}-24~{\Sigma_{2}}^{4} ~\Sigma_{4}-8~{\Sigma_{2}}^{3} ~\Sigma_{6}+44~{\Sigma_{4}}^{2} ~{\Sigma_{2}}^{2}\big),\\
E_3&=7560 \big(6~\mathcal{E}_{12}-4~\mathcal{E}_{2} ~{\Sigma_{10}}+8~\mathcal{E}_{2} ~{\Sigma_{2}}^{5}-19~\mathcal{E}_{4} ~\Sigma_{8}-12~\mathcal{E}_{4} ~{\Sigma_{2}}^{4}+27~\mathcal{E}_{4} ~{\Sigma_{4}}^{2}+68~\mathcal{E}_{6} ~\Sigma_{6}-8~\mathcal{E}_{6} ~{\Sigma_{2}}^{3}\\
&-38~\mathcal{E}_{8} ~\Sigma_{4}+52~\mathcal{E}_{8} ~{\Sigma_{2}}^{2}-20~\mathcal{E}_{10} ~\Sigma_{2}
+26~\mathcal{E}_{2} ~\Sigma_{2} ~\Sigma_{8}+22~\mathcal{E}_{2} ~\Sigma_{2} ~{\Sigma_{4}}^{2}-20~\mathcal{E}_{2} ~\Sigma_{4} ~\Sigma_{6}-8~\mathcal{E}_{2} ~\Sigma_{6} ~{\Sigma_{2}}^{2}\\
&-24~\mathcal{E}_{2} ~\Sigma_{4} ~{\Sigma_{2}}^{3}-40~\mathcal{E}_{4} ~\Sigma_{2} ~\Sigma_{6}+44~\mathcal{E}_{4} ~\Sigma_{4} ~{\Sigma_{2}}^{2}-60~\mathcal{E}_{6} ~\Sigma_{2} ~\Sigma_{4}\big)~,\\
v_3&=
1890 \big(16 ~{\Sigma_2}^5 ~\mathcal{V}_{1,1}-48 ~{\Sigma_2}^4 ~\mathcal{V}_{1,3}-48
  ~\Sigma_4 ~{\Sigma_2}^3 ~\mathcal{V}_{1,1}-32 ~{\Sigma_2}^3 ~\mathcal{V}_{1,5}-16 ~{\Sigma_2}^3
   ~\mathcal{V}_{3,3}-16~\Sigma_6 ~{\Sigma_2}^2 ~\mathcal{V}_{1,1}\\
   &+176~\Sigma_4 ~{\Sigma_2}^2
   ~\mathcal{V}_{1,3}+208 ~{\Sigma_2}^2 ~\mathcal{V}_{1,7}+208 ~{\Sigma_2}^2 ~\mathcal{V}_{3,5}+44
   ~{\Sigma_4}^2~\Sigma_2 ~\mathcal{V}_{1,1}+52~\Sigma_8~\Sigma_2 ~\mathcal{V}_{1,1}-160~\Sigma_6
  ~\Sigma_2 ~\mathcal{V}_{1,3}\\
  &-240~\Sigma_4~\Sigma_2 ~\mathcal{V}_{1,5}-80~\Sigma_2
   ~\mathcal{V}_{1,9}-120~\Sigma_4~\Sigma_2 ~\mathcal{V}_{3,3}-80~\Sigma_2 ~\mathcal{V}_{3,7}-40
  ~\Sigma_2 ~\mathcal{V}_{5,5}-40~\Sigma_4~\Sigma_6 ~\mathcal{V}_{1,1}-8~\Sigma_{10}
   ~\mathcal{V}_{1,1}\\
   &+108 ~{\Sigma_4}^2 ~\mathcal{V}_{1,3}-76~\Sigma_8 ~\mathcal{V}_{1,3}+272
  ~\Sigma_6 ~\mathcal{V}_{1,5}-152~\Sigma_4 ~\mathcal{V}_{1,7}+136~\Sigma_6
   ~\mathcal{V}_{3,3}-152~\Sigma_4 ~\mathcal{V}_{3,5}+24 ~\mathcal{V}_{1,11}+24
   ~\mathcal{V}_{3,9}\\
   &+24 ~\mathcal{V}_{5,7}+52 ~{\Sigma_2}^4 ~\mathcal{U}_4-80 ~{\Sigma_2}^3
   ~\mathcal{U}_6-116~\Sigma_4 ~{\Sigma_2}^2 ~\mathcal{U}_4-116 ~{\Sigma_2}^2 ~\mathcal{U}_8+24~\Sigma_6~\Sigma_2
   ~\mathcal{U}_4+296~\Sigma_4~\Sigma_2 ~\mathcal{U}_6\\
   &+48~\Sigma_2 ~\mathcal{U}_{10}-5 ~{\Sigma_4}^2
   ~\mathcal{U}_4+45~\Sigma_8 ~\mathcal{U}_4-216~\Sigma_6 ~\mathcal{U}_6+102~\Sigma_4 ~\mathcal{U}_8-18
   ~\mathcal{U}_{12}\big)~.\\
\end{aligned}
\end{equation}

The final formulas below correspond to a number of $q=4$ quartets, for a total of 16 particles in the valence shell
\begin{equation}
\label{n4}
\begin{aligned}
\mathcal{N}_4&=22680(16 ~{\Sigma_2}^8-160 ~{\Sigma_4} ~{\Sigma_2}^6+64 ~{\Sigma_6} ~{\Sigma_2}^5+552 ~{\Sigma_4}^2 ~{\Sigma_2}^4+408 ~{\Sigma_8} ~{\Sigma_2}^4-1216 ~{\Sigma_4} ~{\Sigma_6} ~{\Sigma_2}^3-960 ~{\Sigma_{10}} ~{\Sigma_2}^3\\
&-312 ~{\Sigma_4}^3 ~{\Sigma_2}^2+1504 ~{\Sigma_6}^2 ~{\Sigma_2}^2+360 ~{\Sigma_4} ~{\Sigma_8} ~{\Sigma_2}^2+528 ~{\Sigma_{12}} ~{\Sigma_2}^2-336 ~{\Sigma_4}^2 ~{\Sigma_6} ~{\Sigma_2}-2352 ~{\Sigma_6} ~{\Sigma_8}
   ~{\Sigma_2}\\
   &+2016 ~{\Sigma_4} ~{\Sigma_{10}} ~{\Sigma_2}-288 ~{\Sigma_{14}} ~{\Sigma_2}+321 ~{\Sigma_4}^4+944 ~{\Sigma_4} ~{\Sigma_6}^2+1395 ~{\Sigma_8}^2-1206 ~{\Sigma_4}^2 ~{\Sigma_8}-1056 ~{\Sigma_6} ~{\Sigma_{10}}\\
   &-312 ~{\Sigma_4} ~{\Sigma_{12}}+90 {~{\Sigma_{16}}})~.
   \end{aligned}
\end{equation}
\begin{equation}
\label{e4}
\begin{aligned}
E_4&=362880(16 ~\mathcal{E}_2 ~{\Sigma_2}^7-40 ~\mathcal{E}_4 ~{\Sigma_2}^6+24 ~\mathcal{E}_6 ~{\Sigma_2}^5-120 ~\mathcal{E}_2 ~{\Sigma_4} ~{\Sigma_2}^5+204 ~\mathcal{E}_8 ~{\Sigma_2}^4+276 ~\mathcal{E}_4 ~{\Sigma_4} ~{\Sigma_2}^4\\
&+40 ~\mathcal{E}_2 ~{\Sigma_6} ~{\Sigma_2}^4+276 ~\mathcal{E}_2 ~{\Sigma_4}^2 ~{\Sigma_2}^3-600 ~\mathcal{E}_{10} ~{\Sigma_2}^3-456 ~\mathcal{E}_6 ~{\Sigma_4} ~{\Sigma_2}^3-304 ~\mathcal{E}_4 ~{\Sigma_6} ~{\Sigma_2}^3+204 ~\mathcal{E}_2 ~{\Sigma_8} ~{\Sigma_2}^3\\
&-234 ~\mathcal{E}_4
   ~{\Sigma_4}^2 ~{\Sigma_2}^2+396 ~\mathcal{E}_{12} ~{\Sigma_2}^2+180 ~\mathcal{E}_8 ~{\Sigma_4} ~{\Sigma_2}^2+1128 ~\mathcal{E}_6 ~{\Sigma_6} ~{\Sigma_2}^2-456 ~\mathcal{E}_2 ~{\Sigma_4} ~{\Sigma_6} ~{\Sigma_2}^2+90 ~\mathcal{E}_4 ~{\Sigma_8} ~{\Sigma_2}^2\\
   &-360 ~\mathcal{E}_2 ~{\Sigma_{10}} ~{\Sigma_2}^2-78 ~\mathcal{E}_2 ~{\Sigma_4}^3 ~{\Sigma_2}-126 ~\mathcal{E}_6 ~{\Sigma_4}^2 ~{\Sigma_2}+376 ~\mathcal{E}_2 ~{\Sigma_6}^2 ~{\Sigma_2}-252 ~\mathcal{E}_{14} ~{\Sigma_2}+1260
   ~\mathcal{E}_{10} ~{\Sigma_4} ~{\Sigma_2}\\
   &-1176 ~\mathcal{E}_8 ~{\Sigma_6} ~{\Sigma_2}-168 ~\mathcal{E}_4 ~{\Sigma_4} ~{\Sigma_6} ~{\Sigma_2}-882 ~\mathcal{E}_6 ~{\Sigma_8} ~{\Sigma_2}+90 ~\mathcal{E}_2 ~{\Sigma_4} ~{\Sigma_8} ~{\Sigma_2}+504 ~\mathcal{E}_4 ~{\Sigma_{10}} ~{\Sigma_2}\\
   &+132 ~\mathcal{E}_2 ~{\Sigma_{12}} ~{\Sigma_2}+321 ~\mathcal{E}_4 ~{\Sigma_4}^3-603 ~\mathcal{E}_8 ~{\Sigma_4}^2+236 ~\mathcal{E}_4 ~{\Sigma_6}^2-234 ~\mathcal{E}_{12} ~{\Sigma_4}-42 ~\mathcal{E}_2 ~{\Sigma_4}^2 ~{\Sigma_6}-660
   ~\mathcal{E}_{10} ~{\Sigma_6}\\
   &+708 ~\mathcal{E}_6 ~{\Sigma_4} ~{\Sigma_6}+1395 ~\mathcal{E}_8 ~{\Sigma_8}-603 ~\mathcal{E}_4 ~{\Sigma_4} ~{\Sigma_8}-294 ~\mathcal{E}_2 ~{\Sigma_6} ~{\Sigma_8}-396 ~\mathcal{E}_6 ~{\Sigma_{10}}+252 ~\mathcal{E}_2 ~{\Sigma_4} ~{\Sigma_{10}}\\
   &-78 ~\mathcal{E}_4 ~{\Sigma_{12}}-36 ~\mathcal{E}_2 ~{\Sigma_{14}}+90 ~\mathcal{E}_{16})~.\\
   \end{aligned}
\end{equation}
   \begin{equation}
   \label{v4}
\begin{aligned}
   v_4&=90720 \big(32 \mathcal{V}_{1,1} {\Sigma_2}^7+152 \mathcal{U}_4 {\Sigma_2}^6-160
   \mathcal{V}_{1,3} {\Sigma_2}^6-528 \mathcal{U}_6 {\Sigma_2}^5-240 \Sigma_4 \mathcal{V}_{1,1}
   {\Sigma_2}^5+96 \mathcal{V}_{1,5} {\Sigma_2}^5+48 \mathcal{V}_{3,3} {\Sigma_2}^5\\
   &-876
   \Sigma_4 \mathcal{U}_4 {\Sigma_2}^4+180 \mathcal{U}_8 {\Sigma_2}^4+80 \Sigma_6 \mathcal{V}_{1,1}
   {\Sigma_2}^4+1104 \Sigma_4 \mathcal{V}_{1,3} {\Sigma_2}^4+816 \mathcal{V}_{1,7}
   {\Sigma_2}^4+816 \mathcal{V}_{3,5} {\Sigma_2}^4+464 \Sigma_6 \mathcal{U}_4 {\Sigma_2}^3\\
   &+3120
   \Sigma_4 \mathcal{U}_6 {\Sigma_2}^3+2208 \mathcal{U}_{10} {\Sigma_2}^3+552 {\Sigma_4}^2
   \mathcal{V}_{1,1} {\Sigma_2}^3+408 \Sigma_8 \mathcal{V}_{1,1} {\Sigma_2}^3-1216 \Sigma_6
   \mathcal{V}_{1,3} {\Sigma_2}^3-1824 \Sigma_4 \mathcal{V}_{1,5} {\Sigma_2}^3\\   
   &-2400
   \mathcal{V}_{1,9} {\Sigma_2}^3-912 \Sigma_4 \mathcal{V}_{3,3} {\Sigma_2}^3-2400
   \mathcal{V}_{3,7} {\Sigma_2}^3-1200 \mathcal{V}_{5,5} {\Sigma_2}^3+1062 {\Sigma_4}^2
   \mathcal{U}_4 {\Sigma_2}^2+522 \Sigma_8 \mathcal{U}_4 {\Sigma_2}^2\\
   &-4080 \Sigma_6 \mathcal{U}_6
   {\Sigma_2}^2-4212 \Sigma_4 \mathcal{U}_8 {\Sigma_2}^2-1476 \mathcal{U}_{12} {\Sigma_2}^2-912
   \Sigma_4 \Sigma_6 \mathcal{V}_{1,1} {\Sigma_2}^2-720 \Sigma_{10} \mathcal{V}_{1,1}
   {\Sigma_2}^2-936 {\Sigma_4}^2 \mathcal{V}_{1,3} {\Sigma_2}^2\\
   &+360 \Sigma_8 \mathcal{V}_{1,3}
   {\Sigma_2}^2+4512 \Sigma_6 \mathcal{V}_{1,5} {\Sigma_2}^2+720 \Sigma_4 \mathcal{V}_{1,7}
   {\Sigma_2}^2+1584 \mathcal{V}_{1,11} {\Sigma_2}^2+2256 \Sigma_6 \mathcal{V}_{3,3}
   {\Sigma_2}^2+720 \Sigma_4 \mathcal{V}_{3,5} {\Sigma_2}^2\\
   &+1584 \mathcal{V}_{3,9}
   {\Sigma_2}^2+1584 \mathcal{V}_{5,7} {\Sigma_2}^2-744 \Sigma_4 \Sigma_6 \mathcal{U}_4
   \Sigma_2-1224 \Sigma_{10} \mathcal{U}_4 \Sigma_2-684 {\Sigma_4}^2 \mathcal{U}_6 \Sigma_2+2124
   \Sigma_8 \mathcal{U}_6 \Sigma_2\\
   &+7704 \Sigma_6 \mathcal{U}_8 \Sigma_2-5328 \Sigma_4 \mathcal{U}_{10}
   \Sigma_2+1080 \mathcal{U}_{14} \Sigma_2-156 {\Sigma_4}^3 \mathcal{V}_{1,1} \Sigma_2+752
   {\Sigma_6}^2 \mathcal{V}_{1,1} \Sigma_2+180 \Sigma_4 \Sigma_8 \mathcal{V}_{1,1} \Sigma_2\\
   &+264
   \Sigma_{12} \mathcal{V}_{1,1} \Sigma_2-672 \Sigma_4 \Sigma_6 \mathcal{V}_{1,3}
   \Sigma_2+2016 \Sigma_{10} \mathcal{V}_{1,3} \Sigma_2-504 {\Sigma_4}^2 \mathcal{V}_{1,5}
   \Sigma_2-3528 \Sigma_8 \mathcal{V}_{1,5} \Sigma_2-4704 \Sigma_6 \mathcal{V}_{1,7}
   \Sigma_2\\
   &+5040 \Sigma_4 \mathcal{V}_{1,9} \Sigma_2-1008 \mathcal{V}_{1,13} \Sigma_2-252
   {\Sigma_4}^2 \mathcal{V}_{3,3} \Sigma_2-1764 \Sigma_8 \mathcal{V}_{3,3} \Sigma_2-4704
   \Sigma_6 \mathcal{V}_{3,5} \Sigma_2+5040 \Sigma_4 \mathcal{V}_{3,7} \Sigma_2\\
   &-1008
   \mathcal{V}_{3,11} \Sigma_2+2520 \Sigma_4 \mathcal{V}_{5,5} \Sigma_2-1008 \mathcal{V}_{5,9}
   \Sigma_2-504 \mathcal{V}_{7,7} \Sigma_2-399 {\Sigma_4}^3 \mathcal{U}_4+140 {\Sigma_6}^2
   \mathcal{U}_4+693 \Sigma_4 \Sigma_8 \mathcal{U}_4\\
   &+210 \Sigma_{12} \mathcal{U}_4-1752 \Sigma_4
   \Sigma_6 \mathcal{U}_6+1800 \Sigma_{10} \mathcal{U}_6+3483 {\Sigma_4}^2 \mathcal{U}_8-7155
   \Sigma_8 \mathcal{U}_8+3120 \Sigma_6 \mathcal{U}_{10}+990 \Sigma_4 \mathcal{U}_{12}-450
   \mathcal{U}_{16}\\
   &-84 {\Sigma_4}^2 \Sigma_6 \mathcal{V}_{1,1}-588 \Sigma_6 \Sigma_8
   \mathcal{V}_{1,1}+504 \Sigma_4 \Sigma_{10} \mathcal{V}_{1,1}-72 \Sigma_{14}
   \mathcal{V}_{1,1}+1284 {\Sigma_4}^3 \mathcal{V}_{1,3}+944 {\Sigma_6}^2
   \mathcal{V}_{1,3}-2412 \Sigma_4 \Sigma_8 \mathcal{V}_{1,3}\\
   &-312 \Sigma_{12}
   \mathcal{V}_{1,3}+2832 \Sigma_4 \Sigma_6 \mathcal{V}_{1,5}-1584 \Sigma_{10}
   \mathcal{V}_{1,5}-2412 {\Sigma_4}^2 \mathcal{V}_{1,7}+5580 \Sigma_8
   \mathcal{V}_{1,7}-2640 \Sigma_6 \mathcal{V}_{1,9}-936 \Sigma_4 \mathcal{V}_{1,11}\\
   &+360
   \mathcal{V}_{1,15}+1416 \Sigma_4 \Sigma_6 \mathcal{V}_{3,3}-792 \Sigma_{10}
   \mathcal{V}_{3,3}-2412 {\Sigma_4}^2 \mathcal{V}_{3,5}+5580 \Sigma_8
   \mathcal{V}_{3,5}-2640 \Sigma_6 \mathcal{V}_{3,7}-936 \Sigma_4 \mathcal{V}_{3,9}\\
   &+360
   \mathcal{V}_{3,13}-1320 \Sigma_6 \mathcal{V}_{5,5}-936 \Sigma_4 \mathcal{V}_{5,7}+360
   \mathcal{V}_{5,11}+360 \mathcal{V}_{7,9}\big)~.
\end{aligned}
\end{equation}

\end{widetext}

\begin{center}
\begin{table}
\caption{Single particle spectrum and mixing amplitudes for the nucleus $^{32}S$, in both particle and hole formalisms.}
\label{tab1}
\begin{tabular}{c c c c c}
\hline \hline
 s.p. state& $\epsilon$ & $x_{p}^{(q=4)}$ & $x_h^{(q=2)}$  &  $x_{p}^{(q=4)}\cdot x_h^{(q=2)}$\\	[0.5ex]
\hline
 $1d_{5/2}$ &-3.926 & 0.291& 0.0457 & 0.0133 \\
 $2s_{1/2}$& -3.208 &  0.260 & 0.0511& 0.0133\\ 
 $1d_{3/2}$ & 2.112 & 0.0317  & 0.420 & 0.0133\\
 
\hline

\end{tabular}
\end{table}
\end{center}

The above formulas may be employed to compute directly the ground state correlations in $N=Z$ nuclei with up to 16 particles in the valence shell. However, by exploiting the particle-hole symmetry, the same expressions may be applied to cases involving a larger number of particles: a system with a  number of quartets $q$ may be mapped to an equivalent system with $N_{lev}-q$ hole-quartets. Let us note shortly that recently the particle-hole formalism has been used to elaborate  an improved approximate treatment of pairing correlations \cite{Duk16}. Here, the starting point is the reformulation of the PBCS condensate in the particle-hole basis. A detailed study of the generalization of the particle-hole approach to quartet correlations was very recently developed \cite{phb}.

In this paper we limit ourselves to confirm that the particle-hole symmetry is manifest in the framework of the QCM, using the above analytical expressions. We consider as testing ground the nucleus $^{32}S$, which contains 4 quartets, or equivalently 2 hole-quartets, in the valence $sd$ shell. We use the same spherical single particle spectrum as in Ref. \cite{San12}, as displayed in Table I, and assume a constant isovector pairing strength $V_{ij}=-24/A$ MeV, with $A=32$ \cite{Che78}. The transition from particles to holes degrees of freedom may be perfomed in the standard way, resulting in the isovector pairing Hamiltonian in the hole representation
\beq
\label{holeH}
\begin{aligned}
\tilde{H}&=\sum_{i=1}^{N_\text{lev}}\left(4\epsilon_i+3V_{ii}\right)+\sum_{i=1}^{N_\text{lev}}\left(-\epsilon_i-\frac{3}{2}V_{ii}\right) \tilde{N}_{i,0}\\
&+\sum_{\tau=0,\pm1}\sum_{i,j=1}^{N_\text{lev}}V_{ij}\tilde{P}^{\dagger}_{i,\tau } \tilde{P}_{ j,\tau}~,
\end{aligned}
\eeq
where the number of holes operator is $\tilde{N}_{i,0}=4-N_{i,0}$ and the pair operators for holes are defined as $\tilde{P}^{\dagger}_{i,\tau }={P}_{i,\tau }$. Similarly to the pairing case, it turns out that the original quartet "condensate" may be related to a hole-quartet "condensate" of inverse amplitudes (see the Appendix A of \cite{phb} for details):

\beq
 \left(\tilde{Q}^\dagger\left(\frac{1}{x}\right)\right)^k |\text{filled shell}\rangle \propto \left(Q^\dagger\left({x}\right)\right)^{N_{lev}-k} |\text{0}\rangle~.
\eeq

We have confirmed this inverse proportionality by first computing the ground state of $^{32}S$ in a $q=4$ description using the expressions (\ref{n4})-(\ref{v4}), and then a $q=2$ hole-quartet description with the formulas of Eq. (\ref{q2}), with the corresponding modifications of Eq. (\ref{holeH}). In both cases we have obtained a correlation energy $E_\text{corr}=10.36$ MeV. The numerical results regarding the mixing amplitudes are presented In Table I with three significant digits, indicating explicitely the inverse proportionality of particle and hole amplitudes.

Let us finally note that, on the numerical side,  as opposed to the standard recurrence relations method where the running times are of the order of a few
minutes \cite{San12}, our timings are more than two orders of magnitude
smaller (having used the same minimization routine of the NAG library).


In conclusion, we introduce in this paper a method of computer aided analytical calculus for many body problems where it is not only possible, 
but also advantageous, to perform  some algorithmic computations symbolically instead of numerically. This approach presents a twofold benefit: on the numerical side, the computational time may be significantly reduced, and on the implementation side the effort is made negligible. Moreover, it may be applied to a wide class of many body models. In this work, we analyzed the particular example of the Quartet Condensation Model which precisely describes the isovector pairing correlations in $N=Z$ nuclei. 
The corresponding analytical formulas can easily be implemented in any programming language.

The extensions of the QCM model to $N>Z$ nuclei and also to the case of isoscalar pairing are currently under consideration from an analitical perspective
and will be presented in future works.

\acknowledgements
We thank the anonymous referee for the valuable suggestion of analyzing the particle-hole symmetry within our approach.

This work was supported by the grants of the Romanian Ministry of Research and Innovation, CNCS - UEFISCDI, PN-
III-P4-ID-PCE-2016-0092, PN-III-P4-ID-PCE-2016-0792, within PNCDI III, and PN-19060101/2019.



\begin{thebibliography}{99}
\bibitem{Haf38} L.R. Hafstad and E. Teller,  Phys. Rev. {\bf 54}, 681 (1938).
\bibitem{Flo63} B.H. Flowers and M. Vujicik,
Nucl. Phys. {\bf 49}, 586 (1963).
\bibitem{Bri66} D.M. Brink, {\it Proceedings of the International School of Physics Enrico Fermi, Varenna Course} {\bf 36}, 247 (1966). 
\bibitem{Ari71} A. Arima and V. Gillet,
Ann. Phys. {\bf 66}, 117 (1971).
\bibitem{Wil77} K. Wildermuth and Y.C. Tang,
{\it A Unified Theory of the Nucleus} (Academic, New York, 1977). 
\bibitem{Gam83} Y.K. Gambhir, P. Ring, and P. Schuck,
Phys. Rev. Lett. {\bf 51}, 1235 (1983).
\bibitem{Del02} D.S. Delion, G.G. Dussel, and R.J. Liotta,
Rom. J. Phys. {\bf 47}, 97 (2002).
\bibitem{Ike68} K. Ikeda, N. Tagikawa, and H. Horiuchi,
Prog. Theor. Phys. Suppl. {\bf 464} (1968).
\bibitem{Fre97} M. Freer, and A.C. Merchant,
J. Phys. {\bf G} 23, 261 (1997).
\bibitem{Fre07} M. Freer (2007), Rep. Progr. Phys. {\bf 70}, 2149 (2007).
\bibitem{Hor04} H. Horiuchi,
Nucl. Phys. A {\bf 731}, 329 (2004).
\bibitem{Fun09} Y. Funaki, H. Horiuchi, W. von Oertzen, G. R\"opke, 
P. Schuck, A. Tohsaki, and T. Yamada,
Phys.  Rev. C {\bf 80}, 064326 (2009).
\bibitem{Toh17} A. Tohsaki, H. Horiuchi, P. Schuck, and G. R\"opke,
Rev. Mod. Phys. {\bf 89}, 011002 (2017).
\bibitem{He14} W.B. He, Y.G. Ma, X.G. Cao, X.Z. Cai, and G.Q. Zhang,
Phys. Rev. Lett. {\bf 113}, 032506 (2014).
\bibitem{Chi16} Y. Chiba, M. Kimura, Y. Taniguchi,
Phys. Rev. C {\bf 93}, 034319 (2016).
\bibitem{Del10}D.S. Delion,
{\it Theory of particle and cluster emission}
(Springer-Verlag, Berlin, 2010).
\bibitem{Var92} K. Varga, R.G. Lovas, and R.J. Liotta,
Phys. Rev. Lett. {\bf 69}, 37 (1992);
Nucl. Phys. {\bf 550}, 421 (1992).
\bibitem{Del04} D.S. Delion, A. Sandulescu, and W. Greiner,
Phys. Rev. C {\bf 69}, 044318 (2004).
\bibitem{Rop98} G. R\"opke, A. Schnell, P. Schuck, and P. Nozieres,
Phys. Rev. Lett. {\bf 80}, 3177 (1998).
\bibitem{Del13} D.S. Delion and R.J. Liotta,
Phys. Rev. C {\bf 87}, 041302(R) (2013).
\bibitem{San12a} N. Sandulescu, D. Negrea, J. Dukelsky, C. W. Johnson, Phys. Rev. C \textbf{85}, 061303(R) (2012).
\bibitem{danielphd} D. Negrea, {\it Proton-neutron correlations in atomic nuclei}, Ph.D. thesis, 
University of Bucharest and University Paris-Sud, 2013, https://tel.archives-ouvertes.fr/ tel-00870588/document.
\bibitem{San12} N. Sandulescu, D. Negrea, C. W. Johnson,
Phys. Rev. C {\bf 86}, 041302(R) (2012).
\bibitem{Neg14} D. Negrea, N. Sandulescu,  Phys. Rev. C \textbf{90}, 024322 (2014).
\bibitem{San14} N Sandulescu et al, J. Phys.: Conf. Ser. \textbf{533}, 012018 (2014).
\bibitem{San15} N. Sandulescu, D. Negrea, D. Gambacurta, Phys. Lett. B, {\bf 751}, 348 (2015).
\bibitem{Neg17} D. Negrea, N. Sandulescu, D. Gambacurta, Prog. Theor. Exp. Phys.  073D05 (2017).
\bibitem{Sam17} M. Sambataro, N. Sandulescu, Eur. Phys. J. A  \textbf{53} 47 (2017)
\bibitem{Neg18} D. Negrea, P. Buganu, D. Gambacurta,  N. Sandulescu, Phys. Rev. C \textbf{98}, 064319 (2018).
\bibitem{Las18} R.-D. Lasseri, J.-P. Ebran, E. Khan, and N. Sandulescu, Phys. Rev. C {\bf 98}, 014310 (2018).
\bibitem{cdb1} 	K. Peeters, hep-th/0701238.
\bibitem{cdb2}	K. Peeters, Journal of Open Source Software, 3(32), 1118 (2018)
\bibitem{cdb3}https://cadabra.science
\bibitem{danielprivate}D. Negrea, private communication.
\bibitem{Duk16} J. Dukelsky, S. Pittel, and C. Esebbag,
Phys. Rev. C {\bf 93}, 034313 (2016).
\bibitem{phb} V.V. Baran, D.S. Delion, 	arXiv:1902.00065 (2019)
\bibitem{Che78}H.-T. Chen, H. Muther, and A. Faessler, Nucl. Phys. A 297, 445 (1978).
\end{thebibliography}
\end{document}